\documentclass[prb,twocolumn]{revtex4-1}

\usepackage{amsmath}
\usepackage{amsfonts}
\usepackage{graphicx}
\usepackage{hyperref}

\begin{document}

\title{The variational-relaxation algorithm for finding quantum bound states}

\author{Daniel V. Schroeder}
\email{dschroeder@weber.edu}
\affiliation{Department of Physics, Weber State University, Ogden, UT 84408-2508}

\begin{abstract}

I describe a simple algorithm for numerically finding the ground state and low-lying excited states of a quantum system. The algorithm is an adaptation of the relaxation method for solving Poisson's equation, and is fundamentally based on the variational principle. It is especially useful for two-dimensional systems with nonseparable potentials, for which simpler techniques are inapplicable yet the computation time is minimal.  (To be published in the American Journal of Physics.)

\end{abstract}

\maketitle

\section{Introduction}

Solving the time-independent Schr\"odinger equation for an arbitrary potential energy function $V(\vec r)$ is difficult.  There are no generally applicable analytical methods.  In one dimension it is straightforward to integrate the equation numerically, starting at one end of the region of interest and working across to the other.  For bound-state problems for which the energy is not known in advance, the integration must be repeated for different energies until the correct boundary condition at the other end is satisfied; this algorithm is called the shooting method.\cite{Giordano, GTC, Newman, NR}

For a nonseparable\cite{nonseparable} potential in two or more dimensions, however, the shooting method does not work because there are boundary conditions that must be satisfied on all sides. 
One can still use matrix methods,\cite{Marsiglio, Harrison, Schmied} but the amount of computation required can be considerable and the diagonalization routines are mysterious black boxes to most students.

This paper describes a numerical method for obtaining the ground state and low-lying excited states of a bound system in any reasonably small number of dimensions.  The algorithm is closely related to the relaxation method\cite{Relaxation, Koonin, Garcia} for solving Poisson's equation, with the complication that the equation being solved depends on the energy, which is not known in advance.  The algorithm does not require any sophisticated background in quantum mechanics or numerical analysis.  It is reasonably intuitive and easy to code.

The following section explains the most basic version of the algorithm, while Sec.~III derives the key formula using the variational method.  Section~IV presents a two-dimensional implementation of the algorithm in Mathematica.  Section~V generalizes the algorithm to find low-lying excited states, and Sec.~VI presents two nontrivial examples.  The last two sections briefly discuss other related algorithms and how such calculations can be incorporated into the undergraduate physics curriculum.

\section{The algorithm}

A standard exercise in computational physics\cite{Relaxation, Koonin, Garcia} is to solve Poisson's equation,
\begin{equation}
\nabla^2\phi(\vec r) = -\rho(\vec r), \label{poisson}
\end{equation}
where $\rho(\vec r)$ is a known function, by the method of \textit{relaxation}: Discretize space with a rectangular grid, start with an arbitrary function $\phi(\vec r)$ that matches the desired boundary conditions, and repeatedly loop over all the grid points that are not on the boundaries, adjusting each $\phi$ value in relation to its nearest neighbors to satisfy a discretized version of Poisson's equation.  To obtain that discretized version, write each term of the Laplacian operator in the form
\begin{equation}
\frac{\partial^2\phi}{\partial x^2} \approx \frac{\phi(\vec r+\delta\hat x)+\phi(\vec r-\delta\hat x)-2\phi(\vec r)}{\delta^2}, \label{SecondDerivApprox}
\end{equation}
where $\delta$ is the grid spacing and $\hat x$ is a unit vector in the $x$ direction. Solving the discretized Poisson equation for $\phi(\vec r)$ then gives the needed formula,
\begin{equation}
\phi_0 = \overline\phi_\textrm{nn} + \frac1{2d} \rho_0 \delta^2, \label{DiscretePoisson}
\end{equation}
where $\phi_0$ and $\rho_0$ are the values of $\phi$ and $\rho$ at $\vec r$ (the current grid location), $d$ is the dimension of space, and $\overline\phi_\textrm{nn}$ is the average of the $\phi$ values at the $2d$ nearest-neighbor grid locations.  As this formula is applied repeatedly at all grid locations, the array of $\phi$ values ``relaxes'' to the desired self-consistent solution of Poisson's equation that matches the fixed boundary conditions, to an accuracy determined by the grid resolution.

What is far less familiar is that this method can be adapted to solve the time-independent Schr\"odinger equation.  To see the correspondence, write Schr\"odinger's equation with only the Laplacian operator term on the left-hand side:
\begin{equation}
\nabla^2\psi(\vec r) = -2\bigl(E-V(\vec r)\bigr)\psi(\vec r),
\end{equation}
where $E$ is the energy eigenvalue, $V(\vec r)$ is the given potential energy function, and I am using natural units in which $\hbar$ and the particle mass are equal to~1.  Discretizing the Laplacian gives a formula of the same form as Eq.~(\ref{DiscretePoisson}),
\begin{equation}
\psi_0 = \overline\psi_\textrm{nn} + \frac1d(E-V_0)\psi_0\delta^2,
\end{equation}
where the subscripts carry the same meanings as in Eq.~(\ref{DiscretePoisson}). The appearance of $\psi_0$ on the right-hand side creates no difficulty at all, because we can solve algebraically for $\psi_0$:
\begin{equation}
\psi_0 = \frac{\overline\psi_\textrm{nn}}{1 - (E-V_0)\delta^2/d}.
\end{equation}
The more pressing question is what to do with $E$, the energy eigenvalue that we do not yet know.  The answer is that we can replace it with the energy expectation value
\begin{equation}
\langle E\rangle = \frac{\langle\psi|H|\psi\rangle}{\langle\psi|\psi\rangle}, \label{averageE}
\end{equation}
where $H = -\frac12\nabla^2 + V(\vec r)$ is the Hamiltonian operator.  We then update this expectation value after each step in the calculation.  (The denominator in Eq.~(\ref{averageE}) is needed because the algorithm does not maintain the normalization of $\psi$.)  As the relaxation process proceeds $\langle E\rangle$ will steadily decrease, and we will eventually obtain a self-consistent solution for the ground-state energy and wave function.

The inner products in Eq.~(\ref{averageE}) are integrals, but we can compute them to sufficient accuracy as ordinary sums over the grid locations.  The denominator is simply
\begin{equation}
\langle\psi|\psi\rangle = \sum_i \psi_i^2 \delta^d, \label{intPsiSquared}
\end{equation}
where the index $i$ runs over all grid locations and I have assumed that $\psi$ is real. Similarly, the potential energy contribution to the numerator is
\begin{equation}
\langle\psi|V|\psi\rangle = \sum_i V_i\psi_i^2 \delta^d. \label{intPsiVPsi}
\end{equation}
To obtain the kinetic energy ($K=-\frac12\nabla^2$) contribution we again discretize the derivatives as in Eq.~(\ref{SecondDerivApprox}), arriving at the expression
\begin{equation}
\langle\psi|K|\psi\rangle = -d\sum_i (\psi_i\overline\psi_\textrm{nn}-\psi_i^2)\delta^{d-2}.\label{intPsiKPsi}
\end{equation}
Each of these inner products must be updated after every change to one of the $\psi_i$ values, but there is no need to evaluate them from scratch.  When we change $\psi_{0,\,\textrm{old}}$ to $\psi_{0,\,\textrm{new}}$, the corresponding changes to the inner products are
\begin{align}
\Delta\langle\psi|\psi\rangle &= (\psi_{0,\,\textrm{new}}^2 -\psi_{0,\,\textrm{old}}^2)\delta^d,\label{intPsiSquaredChange} \\
\Delta\langle\psi|V|\psi\rangle &= V_0(\psi_{0,\,\textrm{new}}^2 -\psi_{0,\,\textrm{old}}^2)\delta^d, \\
\Delta\langle\psi|K|\psi\rangle &= -2d(\psi_{0,\,\textrm{new}}-\psi_{0,\,\textrm{old}})\overline\psi_\textrm{nn}\delta^{d-2} \nonumber \\
& \qquad + d(\psi_{0,\,\textrm{new}}^2 -\psi_{0,\,\textrm{old}}^2)\delta^{d-2},\label{intPsiKPsiChange}
\end{align}
where the factor of 2 in the first term of Eq.~(\ref{intPsiKPsiChange}) arises because there is an identical contribution of this form from the terms in the sum of Eq.~(\ref{intPsiKPsi}) in which $i$ is one of the neighboring grid locations.

The algorithm, then, is as follows:
\begin{enumerate}
\item Discretize space into a rectangular grid, placing the boundaries far enough from the region of interest that the ground-state wave function will be negligible there.
\item Initialize the array of $\psi$ values to represent a smooth, nodeless function such as the ground state of an infinite square well or a harmonic oscillator. All the $\psi$ values on the boundaries should be zero and will remain unchanged.
\item Use Eqs.\ (\ref{intPsiSquared})--(\ref{intPsiKPsi}) to calculate $\langle\psi|\psi\rangle$, $\langle\psi|H|\psi\rangle$, and $\langle E\rangle$ for the initial $\psi$ array.
\item Loop over all interior grid locations, setting the $\psi$ value at each location to
\begin{equation}
\psi_0 = \frac{\overline\psi_\textrm{nn}}{1 - (\langle E\rangle-V_0)\delta^2/d}.
\label{psi0formula}
\end{equation}
Also use Eqs.\ (\ref{intPsiSquaredChange})--(\ref{intPsiKPsiChange}) to compute the changes to $\langle\psi|H|\psi\rangle$ and $\langle\psi|\psi\rangle$ that result from this change to $\psi_0$, and use these quantities to update the value of $\langle E\rangle$ before proceeding to the next grid location.
\item Repeat step 4 until $\langle E\rangle$ and $\psi(\vec r)$ no longer change, within the desired accuracy.
\end{enumerate}

The simplest procedure, as just described, is to update each $\psi$ value ``in place,'' so that a change at one grid location immediately affects the calculation for the next grid location.  In the terminology of relaxation methods, this approach is called the \textit{Gauss-Seidel} algorithm.\cite{Relaxation, Koonin, Garcia}

\section{Variational interpretation}

In the previous section I asserted, but did not prove, that $\langle E\rangle$ will steadily decrease during the relaxation process.  To see why this happens, it is instructive to derive Eq.~(\ref{psi0formula}) using the variational method of quantum mechanics.\cite{Variational} The idea is to treat each local value $\psi_0$ as a parameter on which the function $\psi(\vec r)$ depends, and repeatedly adjust these parameters, one at a time, to minimize the energy expectation value $\langle E\rangle$.  So let us consider how the expression for $\langle E\rangle$ in Eq.~(\ref{averageE}) depends on~$\psi_0$.

Focusing first on the denominator of Eq.~(\ref{averageE}), we discretize the integral as in Eq.~(\ref{intPsiSquared}), but rewrite the sum as
\begin{equation}
\langle\psi|\psi\rangle = \psi_0^2 \delta^d + s,
\label{denominator}
\end{equation}
where $s$ is an abbreviation for the terms in the sum that do not depend on $\psi_0$. Similarly, the discretization of Eqs.~(\ref{intPsiVPsi}) and (\ref{intPsiKPsi}) allows us to write the numerator of Eq.~(\ref{averageE}) as
\begin{equation}
\langle\psi|H|\psi\rangle = -2d\psi_0\overline\psi_\textrm{nn}\delta^{d-2} + d\psi_0^2\delta^{d-2}
 + V_0\psi_0^2\delta^d + h,
\label{numerator}
\end{equation}
where the factor of 2 in the first term is the same as in Eq.~(\ref{intPsiKPsiChange}) and $h$ is an abbreviation for all the terms that do not depend on $\psi_0$.

\begin{figure*}[t!]
\centering
\includegraphics[width=17cm]{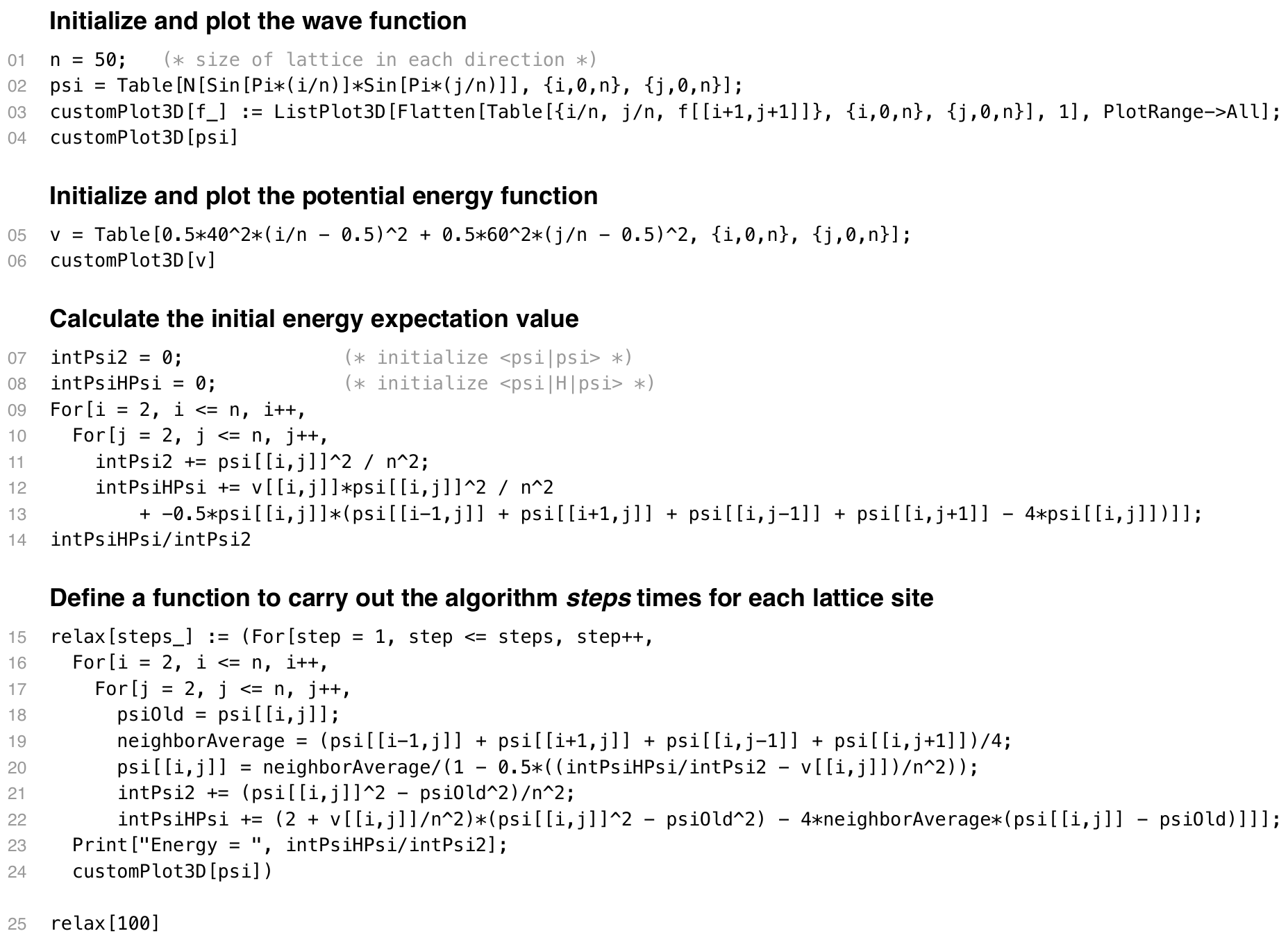}
\caption{Mathematica code to implement the basic variational-relaxation algorithm for a two-dimensional quantum system. Here the potential energy function is for a harmonic oscillator, for which the solutions are known analytically.}
\end{figure*}

Inserting Eqs.~(\ref{denominator}) and (\ref{numerator}) into Eq.~(\ref{averageE}) gives
\begin{equation}
\langle E\rangle = \frac{h - 2d\delta^{d-2}\psi_0\overline\psi_\textrm{nn} + (d\delta^{d-2}{+}V_0\delta^d)\psi_0^2}{s+\delta^d\psi_0^2}, \label{ExplicitAverageE}
\end{equation}
where I have written the $h$ and $s$ terms first because they are larger than the others by a factor on the order of the total number of lattice points.\cite{WhatIfNegativeV} 
We are looking for the value of $\psi_0$ that minimizes this expression. Differentiating with respect to $\psi_0$ and setting the result equal to 0 gives a complicated equation, but in the limit of a large lattice it is a valid approximation to keep only the leading terms in $s$ and $h$. With that approximation, after some algebra, the extremization condition reduces to
\begin{equation}
\psi_0 = \frac{\overline\psi_\textrm{nn}}{1 - [(h/s)-V_0]\delta^2/d}.
\end{equation}
The ratio $h/s$ is equal to $\langle E\rangle$ in the limit of an infinite lattice, so this result is effectively the same as Eq.~(\ref{psi0formula}). By similarly focusing on the leading nontrivial terms in powers of $s$ and $h$ it is straightforward to show that this extremum is a minimum, if the lattice spacing $\delta$ is sufficiently small.

We can therefore be confident that each step of the algorithm will reduce the value of $\langle E\rangle$. This result suggests, but does not prove, that the algorithm will converge to the system's ground state. In fact every energy eigenfunction is a stationary point of the energy functional,\cite{Variational} so there can be situations in which the algorithm converges (or almost converges) to an excited state instead of the ground state. But the excited states are unstable to small perturbations, and they can be avoided entirely by choosing an initial wave function that is sufficiently similar to the ground state. Once the algorithm brings $\langle E\rangle$ below every excited-state energy, the ground state is the only possible result after sufficiently many iterations.\cite{TrivialNote}

\section{An implementation}

Figure 1 shows a basic implementation of the variational-relaxation algorithm in Mathematica,\cite{Mathematica} for a two-dimensional potential well. Translating this example to other computer languages should be straightforward.

The first four lines of the code define the resolution of the lattice (here $50\times50$), initialize the wave function to the ground state of an infinite square well, and then plot the initial wave function using a custom plotting function that maps the array of lattice points to a square in the $xy$ plane extending from 0 to 1 in each direction. Notice that the array size is one element larger in each dimension than the nominal lattice size ($51\times51$ in this case), so that the edges can be mapped to exactly 0 and~1, where the wave function will be held fixed throughout the calculation. Notice also that an offset of 1 is required when indexing into the array, because Mathematica array indices start at 1 rather than 0.

Lines 5 and 6 define and plot an array of values to represent the potential energy function. Here, for testing purposes, this function is a harmonic oscillator potential with a classical angular frequency of 40 (in natural units) in the $x$ direction and 60 in the $y$ direction. The rest of the code is sufficiently versatile, however, that almost any potential energy function can be used, as long as its discrete representation is reasonably accurate.

Lines 7--13 compute the inner products $\langle\psi|\psi\rangle$ and $\langle\psi|H|\psi\rangle$ according to Eqs.~(\ref{intPsiSquared}) through (\ref{intPsiKPsi}). Because the width of the two-dimensional space is one unit, the lattice spacing $\delta$ is the reciprocal of the lattice size (1/50). Line 14 displays the initial value of $\langle E\rangle$.

The algorithm itself is implemented in the \texttt{relax} function (lines 15--24), whose argument is the number of times to iterate the algorithm for each lattice site. For each iteration step we loop over all the lattice sites and for each site, save the old wave function value, calculate the new value from Eq.~(\ref{psi0formula}), and update the inner products $\langle\psi|\psi\rangle$ and $\langle\psi|H|\psi\rangle$ using Eqs.~(\ref{intPsiSquaredChange})--(\ref{intPsiKPsiChange}). When everything is finished we display the final value of $\langle E\rangle$ and then plot the final wave function. To actually execute this function we type something like \texttt{relax[100]} for 100 iteration steps. We can do this repeatedly, checking the results for convergence.

For this harmonic oscillator example using a $50\times50$ lattice, 100 iteration steps results in an energy value of 49.97, within less than 0.1\% of the analytically known value of 50 (that is, $\hbar/2$ times the sum of the $x$ and $y$ frequencies). After another 100 steps the energy converges to 49.94, slightly below the analytical value due to the lattice discretization.  The calculated wave function has the familiar Gaussian shape. On a typical laptop computer, Mathematica can execute 100 iteration steps for a $50\times50$ lattice in just a few seconds. This execution speed, along with the brevity of the code, brings two-dimensional calculations of this type well within the reach of a typical undergraduate homework assignment.

\section{Extensions}

An easy trick for speeding up the algorithm is to use \textit{over-relaxation},\cite{Relaxation, Koonin, Garcia} in which we try to anticipate subsequent iterations by ``stretching'' each change to a $\psi$ value by a factor between 1 and 2. If we call the value of expression (\ref{psi0formula}) $\psi_{0,\textrm{nominal}}$, then the formula to update $\psi_0$ becomes
\begin{equation}
\psi_{0,\,\textrm{new}} = \psi_{0,\,\textrm{old}} + \omega (\psi_{0,\,\textrm{nominal}}-\psi_{0,\,\textrm{old}}),
\end{equation}
where the ``stretch factor'' $\omega$ is called the over-relaxation parameter. Figure~2 shows how the rate of convergence depends on $\omega$, for the two-dimensional harmonic oscillator example described in the preceding section.

\begin{figure}[t]
\centering
\includegraphics[width=8.0cm]{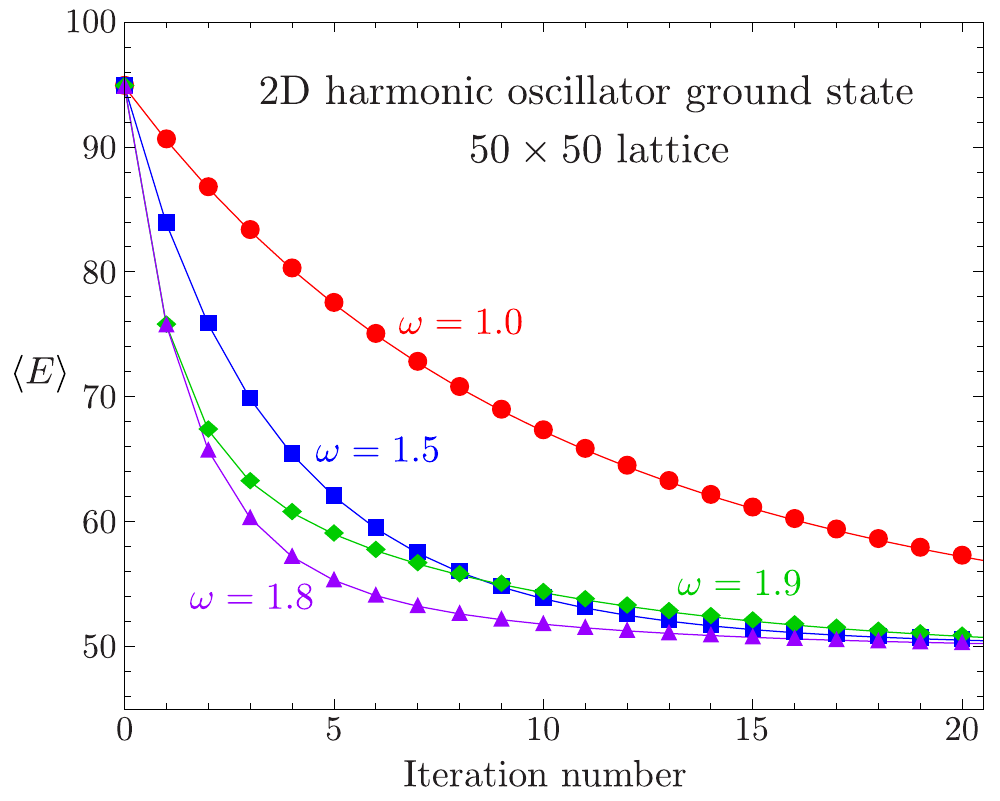}
\caption{The energy expectation value $\langle E\rangle$ as a function of the iteration number, for the two-dimensional harmonic oscillator example used in Sec.~IV. The different data sets are for different values of the over-relaxation parameter $\omega$. The basic algorithm without over-relaxation corresponds to $\omega=1$. In this example, with a lattice size of $50\times50$, the optimum $\omega$ is about~1.8.}
\end{figure}

After finding the ground state of a particular system, we can go on to find its first excited state with only a minor modification of the algorithm. The idea is the same as with other variational solutions,\cite{Variational} namely, to restrict the trial wave function to be orthogonal to the ground state. To do this, we periodically project out any contribution of the ground state to the trial function during the course of the calculation. More explicitly, the procedure is as follows:

\begin{enumerate}
\item Normalize and save the just-determined ground-state wave function as $\psi_\textrm{gs}$.
\item Initialize a new trial wave function $\psi$ that crudely resembles the first excited state, with a single node. The first excited state of an infinite square well or a harmonic oscillator would be a reasonable choice. It may be necessary to try different orientations for the node of this function.
\item Proceed as in the basic algorithm described in Sec.~II, but after each loop through all the grid locations, calculate the $\psi_\textrm{gs}$ component of the trial function as the inner product
\begin{equation}
\langle\psi_\textrm{gs}|\psi\rangle = \sum_i \psi_{\textrm{gs},i}\psi_i\delta^d.
\end{equation}
Multiply this inner product by $\psi_\textrm{gs}$ and subtract the result from $\psi$ (point by point). Then recalculate the inner products $\langle\psi|\psi\rangle$ and $\langle\psi|H|\psi\rangle$ before proceeding to the next iteration.
\end{enumerate}

The orientation of the initial state's node matters because we want it to resemble the first excited state more than the second. For example, the first excited state of the anisotropic harmonic oscillator potential used in Sec.~IV has a node line parallel to the $y$~axis, so a good choice for the initial state would be $\sin(2\pi x)\sin(\pi y)$, rather than the orthogonal state $\sin(\pi x)\sin(2\pi y)$. If the latter state is used the algorithm will become stuck, for a rather long time, on the second excited state (with energy 110) before finally converging to the first excited state (with energy 90).

After finding the first excited state we can find the second excited state in a similar way, this time projecting out both the ground-state contribution and the first-excited-state contribution after each loop through all the grid locations. We could then go on to find the third excited state and so on, but if many states are needed it may be easier to use matrix methods.\cite{Marsiglio, Harrison, Schmied}

\section{Examples}

\begin{figure}[b!]
\centering
\includegraphics[width=8.5cm]{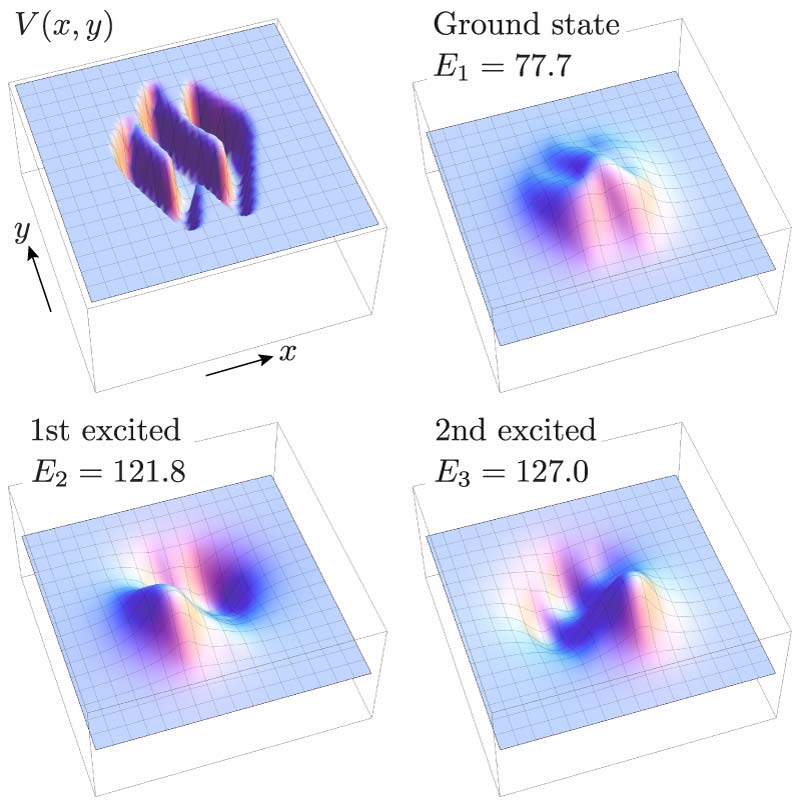}
\caption{A ``Flaming W'' potential energy well (upper left) and the three lowest-energy wave functions and corresponding energies for a particle trapped in this well.  The potential energy is zero inside the W and $+200$ (in natural units) in the flat area surrounding it.  The grid resolution is $64\times64$.}
\end{figure}

To illustrate the versatility of the variational-relaxation algorithm, Fig.~3 shows results for an intricate but contrived potential energy well based on an image of the Weber State University ``Flaming W'' logo.\cite{FlamingW}  As before, the units are such that $\hbar$, the particle mass, and the width of the square grid region are all equal to~1.  In these units the sine-wave starting function (that is, the ground state of an infinitely deep two-dimensional box of this size) has a kinetic energy of $\pi^2\approx10$, so the well depth of 200 is substantially larger than this natural energy scale. All three of the states shown are bound, with energies less than 200. As expected, the ground-state wave function spreads to fill the oddly shaped potential well, but is peaked near the center. The two lowest excited states are relatively close together in energy, with nodal curves that are roughly orthogonal to each other.

\begin{figure}[b!]
\centering
\includegraphics[width=8.5cm]{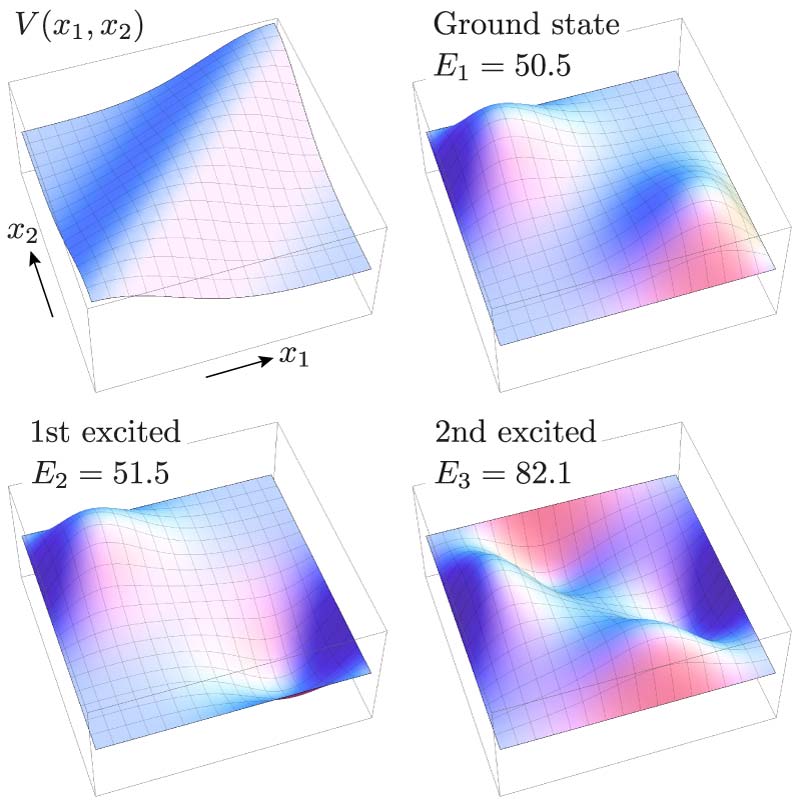}
\caption{The interparticle potential energy (upper left) and the three lowest-energy wave functions and corresponding energies for a pair of equal-mass but distinguishable particles trapped in a one-dimensional infinite square well, repelling each other according to Eq.~(\ref{GaussianRepulsion}). The grid resolution is $50\times50$ and the maximum potential energy is 80 in natural units.}
\end{figure}

For a second example, note that the Schr\"odinger equation for a single particle in two dimensions is mathematically equivalent to that for two equal-mass particles in one dimension. We can therefore adapt our results to the latter system by renaming $(x,y)\rightarrow(x_1,x_2)$. Consider, then, a pair of equal-mass (but distinguishable) particles trapped in a one-dimensional infinite square well, exerting a repulsive force on each other.\cite{cmAndRelative} A smooth potential for modeling such a force is a Gaussian,\cite{Maestri}
\begin{equation}
V(x_1,x_2) = V_{\max} e^{-(x_1-x_2)^2/a^2}, \label{GaussianRepulsion}
\end{equation}
and for illustration purposes I will take $V_{\max}=80$ and $a=0.4$ in natural units. This potential and its three lowest-energy stationary states are shown in Fig.~4. Interpreting these two-particle wave function plots takes a little practice; for example, the two peaks in the ground-state wave function correspond not to the two particles, but rather to two equally probable configurations for both particles, one with particle~1 near $x=0$ and particle~2 near $x=1$, and the other with the particles interchanged. This is an ``entangled'' state, because a measurement of one particle's position changes the probability distribution for the other particle's position. Notice that the first excited state, with a node along $x_1=x_2$, has an almost identical probability density and only slightly more energy, as is typical of double-well potentials. In contrast, the second excited state tends to put one particle or the other near the middle of the well and has considerably more energy.

These two examples are merely meant to suggest the wide range of possible uses of the variational-relaxation algorithm.  The algorithm should be applicable to real-world systems such as quantum dots\cite{Harrison, QuantumDots} and other nano-structures that can be modeled as two-dimensional or three-dimensional potential wells.  For a system of two particles in one dimension, one could investigate other interaction potentials, repulsive or attractive, as well as other confining potentials.

\section{Related algorithms}

The algorithm described in this paper cannot possibly be new, because it is such a minor adaptation of the familiar relaxation algorithm for Poisson's equation. However, I have been unable to find a published description of it.\cite{NotInNR, Ames}

Giordano and Nakanishi\cite{GiordanoMonteCarlo} describe a closely related algorithm that also uses a rectangular lattice and the variational principle, but takes a Monte Carlo approach. Instead of looping over all lattice points in order, they choose successive lattice points at random. And instead of computing the value of $\psi_0$ that minimizes $\langle E\rangle$ using Eq.~(\ref{psi0formula}), they consider a random change to $\psi_0$, compute the effect of this change on $\langle E\rangle$, and then accept the change if $\langle E\rangle$ will decrease but reject it if $\langle E\rangle$ would increase. This Monte Carlo approach inspired the algorithm described in this paper. However, the Monte Carlo algorithm is much less efficient, at least when fixed, uniform distributions are used for the random numbers.

Koonin and Meredith\cite{KooninFakeTime} describe an alternative algorithm that evolves an initial trial function in imaginary time, according to the Schr\"odinger-like equation
\begin{equation}
\label{FakeTimeTDSE}
\frac{\partial\psi}{\partial\tau} = -H\psi,
\end{equation}
whose formal solution is
\begin{equation}
\psi(\tau) = e^{-H\tau}\psi(0).
\end{equation}
If we imagine expanding $\psi(0)$ as a linear combination of eigenfunctions of the Hamiltonian $H$, then we see that the ground-state term in the expansion decreases the most slowly (or grows the most rapidly if its energy is negative), so eventually this evolution in ``fake time'' will suppress all the remaining terms in the expansion and yield a good approximation to the ground state.\cite{FakeTimeNote} An advantage of the imaginary-time approach is that its validity rests on the fundamental argument just given, rather than on the more subtle variational principle.

The speed and coding complexity of an imaginary-time algorithm depend on the specific method used for the imaginary-time evolution.  Koonin and Meredith use a basic first-order forward-time Euler integration, resulting in an algorithm that is just as easy to code as the variational-relaxation algorithm, and that requires about the same execution time as the latter without over-relaxation. Their algorithm is therefore a strong candidate for use in an undergraduate course, especially if students are more familiar with time-evolution algorithms than with relaxation algorithms (and if teaching relaxation algorithms is not a course goal).

Faster imaginary-time algorithms also exist, but may be too sophisticated for many educational settings. Simply switching to a centered-difference approximation for the time derivative, which is quite effective for the actual time-dependent Schr\"odinger equation,\cite{Visscher} yields an algorithm that is unstable no matter how small the time step.\cite{AmesUnstable} Implicit algorithms\cite{Implicit} would solve the stability problem, but these require working with large matrices.  One reviewer of early drafts of this paper strongly recommends an imaginary-time adaptation of the split-operator algorithm described in Sec.~16.6 of Ref.~\onlinecite{GTC}, which uses a fast Fourier transform to switch back and forth between position space and momentum space during each time step.

\section{Classroom use}

Students in a computational physics course should be able to code the variational-relaxation algorithm themselves, perhaps after practicing on the ordinary relaxation algorithm for Poisson's or Laplace's equation.  Coding the algorithm in just one spatial dimension can also be a good warm-up exercise, keeping in mind that it is usually easier to solve one-dimensional problems by the shooting method.

In an upper-division undergraduate course in quantum mechanics, it may be better to provide students with the basic code shown in Fig.~1 (or its equivalent in whatever programming language they will use).  Typing the code into the computer gives students a chance to think about each computational step.  After verifying that the algorithm works for a familiar example such as the two-dimensional harmonic oscillator, students can be asked to modify it to handle other potential energy functions, over-relaxation, and low-lying excited states.

\parskip=0pt

Even in a lower-division ``modern physics'' course or the equivalent, I think there is some benefit in showing students that the two-dimensional time-independent Schr\"odinger equation, for an arbitrary potential energy function, can be solved.  For the benefit of students and others who are not ready to code the algorithm themselves, and for anyone who wishes to quickly explore some nontrivial two-dimensional stationary states, the electronic supplement\cite{supplement} to this paper provides a JavaScript implementation of the algorithm with a graphical user interface, runnable in any modern web browser.

In any of these settings, and in any other physics course, introducing general-purpose numerical algorithms can help put the focus on the laws of physics themselves, avoiding an over-emphasis on idealized problems and specialized analytical tricks.

\newpage\raggedbottom

\vglue1ex

\begin{acknowledgments}
I am grateful to Nicholas Giordano, Hisao Nakanishi, and Saul Teukolsky for helpful correspondence, to Colin Inglefield for bringing Ref.~\onlinecite{Harrison} to my attention, to the anonymous reviewers for many constructive suggestions, and to Weber State University for providing a sabbatical leave that facilitated this work.
\end{acknowledgments}


\begin{thebibliography}{99}

\bibitem{Giordano} N. J. Giordano and H. Nakanishi, \textit{Computational Physics}, 2nd ed. (Pearson Prentice Hall, Upper Saddle River, NJ, 2006), Sec.~10.2.

\bibitem{GTC} H. Gould, J. Tobochnik, and W. Christian, \textit{An Introduction to Computer Simulation Methods}, 3rd ed. (Pearson, San Francisco, 2007), \url{http://www.compadre.org/osp/items/detail.cfm?ID=7375}, Sec.~16.3.

\bibitem{Newman} M. Newman, \textit{Computational Physics}, revised edition (CreateSpace, Seattle, 2013), Sec.~8.6.

\bibitem{NR} W. H. Press, S. A. Teukolsky, W. T. Vetterling, and B. P. Flannery, \textit{Numerical Recipes}, 3rd ed. (Cambridge University Press, Cambridge, 2007), Sec.~18.1.

\bibitem{nonseparable}  When the potential energy function can be written as a sum of one-dimensional potentials, e.g., $V(x)+V(y)$, the time-independent Schr\"odinger equation can be solved by separation of variables, reducing its solution to the one-dimensional case.  However, a generic multidimensional potential does not have this property.

\bibitem{Marsiglio} F. Marsiglio, ``The harmonic oscillator in quantum mechanics: A third way,'' Am. J. Phys. \textbf{77}(3), 253--258 (2009); R. L. Pavelich and F. Marsiglio, ``Calculation of 2D electronic band structure using matrix mechanics,'' Am. J. Phys. \textbf{84}(12), 924--935 (2016).

\bibitem{Harrison} P. Harrison and A. Valavanis, \textit{Quantum Wells, Wires and Dots}, 4th ed. (John Wiley \& Sons, Chichester, UK, 2016).

\bibitem{Schmied} R. Schmied, \textit{Introduction to Computational Quantum Mechanics} (unpublished lecture notes, 2016), \url{http://atom.physik.unibas.ch/teaching/CompQM.pdf}.

\bibitem{Relaxation} See Giordano and Nakanishi, Ref.~\onlinecite{Giordano}, Sec.~5.2; Gould, Tobochnik, and Christian, Ref.~\onlinecite{GTC}, Sec.~10.5; Newman, Ref.~\onlinecite{Newman}, Secs.~9.1--9.2; and Press, \textit{et al.}, Ref.~\onlinecite{NR}, Sec.~20.5.

\bibitem{Koonin} S. E. Koonin and D. C. Meredith, \textit{Computational Physics: Fortran Version} (Addison-Wesley, Reading, MA, 1990), Sec.~6.2.

\bibitem{Garcia} A. L. Garcia, \textit{Numerical Methods for Physics}, 2nd ed., revised (CreateSpace, Seattle, 2015), Sec.~8.1.

\bibitem{Variational} The variational method is discussed in most quantum mechanics textbooks. Especially thorough treatments can be found in C. Cohen-Tannoudji, B. Diu, and F. Lalo\"e, \textit{Quantum Mechanics} (John Wiley \& Sons, New York, 1977), Vol.~II, Complement E$_\textrm{XI}$; E. Merzbacher, \textit{Quantum Mechanics}, 3rd ed. (John Wiley \& Sons, New York, 1998), Sec.~8.1; and R. Shankar, \textit{Principles of Quantum Mechanics}, 2nd ed. (Springer, New York, 1994), Sec.~16.1.  Each of these texts shows more generally that every discrete stationary state is an extremum of the energy functional $\langle E\rangle$.

\bibitem{WhatIfNegativeV} The claim that $h$ is much larger than the other terms in the numerator of Eq.~(\ref{ExplicitAverageE}) is valid if $V(\vec r)$ is always positive. The potential energy can always be shifted so that this condition holds.

\bibitem{TrivialNote} The trivial function $\psi(x)=0$ is also a solution of Eq.~(\ref{psi0formula}), but is unstable under small perturbations so the algorithm will never converge to this solution.

\bibitem{Mathematica} Wolfram Mathematica, \url{http://www.wolfram.com/mathematica/}.

\bibitem{FlamingW} Weber State University Logos, \url{http://departments.weber.edu/marcomm/logodownloads/}.

\bibitem{cmAndRelative} When the two-particle potential consists only of an interaction term of the form $V(x_1-x_2)$, the Schr\"odinger equation is separable in terms of center-of-mass and relative coordinates.  Adding a confining potential, such as the infinite square well used here, makes the problem irreducibly two-dimensional.

\bibitem{Maestri} For simulation results of scattering interactions between two particles in one dimension interacting via a Gaussian potential, see J. J. V. Maestri, R. H. Landau, and M. J. P\'aez, ``Two-particle Schr\"odinger equation animations of wave packet-wave packet scattering,'' Am. J. Phys. \textbf{68}(12), 1113--1119 (2000).

\bibitem{QuantumDots} For other approaches to analyzing quantum dots, see Z. M. Schultz and J. M. Essick, ``Investigation of exciton ground state in quantum dots via Hamiltonian diagonalization method,'' Am. J. Phys. \textbf{76}(3), 241--249 (2008); B. J. Riel, ``An introduction to self-assembled quantum dots,'' Am. J. Phys. \textbf{76}(8), 750--757 (2008); D. Zhou and A. Lorke, ``Wave functions of elliptical quantum dots in a magnetic field,'' Am. J. Phys. \textbf{83}(3), 205--209 (2015).

\bibitem{NotInNR} Section 18.3 of \textit{Numerical Recipes}, Ref.~\onlinecite{NR}, presents a general relaxation algorithm for systems of ordinary differential equations, and Sec.~8.0.1 shows how to treat eigenvalue problems by adding another equation to the system. For the special case of the one-dimensional time-independent Schr\"odinger equation the approach taken in this paper is much simpler. Reference~\onlinecite{NR} does not discuss eigenvalue problems for partial differential equations.

\bibitem{Ames} See, for example, W. F. Ames, \textit{Numerical Methods for Partial Differential Equations}, 2nd ed. (Academic Press, New York, 1977). This standard reference includes a section on eigenvalue problems for partial differential equations, but does not describe any algorithm that I can recognize as equivalent to the one in this paper.

\bibitem{GiordanoMonteCarlo} Giordano and Nakanishi, Ref.~\onlinecite{Giordano}, Sec.~10.4.

\bibitem{KooninFakeTime} Koonin and Meredith, Ref.~\onlinecite{Koonin}, Sec.~7.4.

\bibitem{FakeTimeNote} We can interpret the imaginary time parameter $\tau$ as an inverse temperature, and the exponential factor $e^{-E\tau}$ as a Boltzmann probability factor in the canonical ensemble. Then the limit $\tau\rightarrow\infty$ corresponds to lowering the reservoir temperature to zero, putting the system into its ground state.

\bibitem{Visscher} P. B. Visscher, ``A fast explicit algorithm for the time-dependent Schr\"odinger equation,'' Computers in Physics \textbf{5} (6), 596--598 (1991). This algorithm is also described in Ref.~\onlinecite{Giordano}, Sec.~10.5, and Ref.~\onlinecite{GTC}, Sec.~16.5; I learned it from T. A. Moore in 1982.

\bibitem{AmesUnstable} See Ames, Ref.~\onlinecite{Ames}, Sec.~2-4. The argument given there is easily adapted to Eq.~(\ref{FakeTimeTDSE}).

\bibitem{Implicit} See Koonin and Meredith, Ref.~\onlinecite{Koonin}, Secs. 7.2--7.3; Garcia, Ref.~\onlinecite{Garcia}, Chap.~9; or Press \textit{et al.}, Ref.~\onlinecite{NR}, Secs.~20.2--20.3.

\bibitem{supplement} See the ``Quantum Bound States in Two Dimensions'' web app at \url{http://physics.weber.edu/schroeder/software/BoundStates2D.html}.

\end{thebibliography}
\end{document}